%% file: Paper.tex
\documentclass[aps,showpacs,twocolumn,twoside,amsmath,amssymb,superscriptaddress,prc]{revtex4-2}
\usepackage[english]{babel}
\usepackage{multirow}
\usepackage{tikz}
\usepackage{epstopdf}
\usepackage[percent]{overpic}  
\usepackage{scrextend}  
\usepackage{xcolor,xspace}
\usepackage[ulem=normalem]{changes}
\usepackage{hyperref}
\hypersetup{colorlinks=True,urlcolor=blue,linkcolor=blue,citecolor=blue,filecolor=black}

\colorlet{Changes@Color}{red}  
\usepackage{amsmath}
\usepackage{dcolumn,color,footnote,bm,braket}
\usepackage{url,longtable,tabularx}

\usepackage{fancybox}
\usetikzlibrary{matrix}

\begin{document}

\title{
Odd nuclei and quasiparticle excitations within the Barcelona Catania Paris Madrid energy
density functional
}

\author{S.~A.~Giuliani}
\email{samuel.giuliani@uam.es}
\affiliation{Departamento de F\'\i sica Te\'orica and CIAFF, Universidad
Aut\'onoma de Madrid, E-28049 Madrid, Spain}

\affiliation{Department of Physics, Faculty of Engineering and Physical Sciences, 
University of Surrey, Guildford, Surrey GU2 7XH, United Kingdom.}

\author{L.~M.~Robledo}
\email{luis.robledo@uam.es}
\affiliation{Departamento de F\'\i sica Te\'orica and CIAFF, Universidad
Aut\'onoma de Madrid, E-28049 Madrid, Spain}

\affiliation{Center for Computational Simulation,
Universidad Polit\'ecnica de Madrid,
Campus de Montegancedo, Bohadilla del Monte, E-28660-Madrid, Spain}

\date{\today}

\begin{abstract}
An extension of the Barcelona Catania Paris Madrid (BCPM) energy density
functional is proposed to deal with odd-mass systems as well as
multiquasiparticle excitations. The extension is based on the assumption that
the equal filling approximation (EFA) is a valid alternative to the traditional
full blocking procedure of the Hartree-Fock-Bogoliubov method. The assumption is
supported by the excellent agreement between full blocking and EFA calculations
obtained with different parametrizations of the Gogny interaction. The EFA
augmented BCPM functional is used to compute low energy excitation spectra of
selected nuclei in different regions of the nuclear chart, and high-$K$ isomers
in $^{178}$Hf. We show that BCPM predictions are in good agreement with Gogny
D1M results and experimental data.
\end{abstract}

\maketitle


\section{Introduction}


Nuclei with an odd number of protons or neutrons are an archetypical example of
spin-polarized fermionic systems. Due to the presence of an unpaired nucleon,
their theoretical description using mean-field models poses additional
challenges compared to even-even systems. This is because the short-range
nature of pairing interaction couple the nucleons in Cooper pairs in order to
maximize their spatial overlap, simplifying the
description of nuclei with an even number of protons and neutrons. Conversely,
odd-$A$ nuclei break time-reversal symmetry, hence removing Kramers
degeneracy in single-particle levels and requiring the inclusion of
energy density functionals (EDF) time-odd fields. This results in an increased
computational cost, which has traditionally limited the number of calculations
devoted to odd-$A$ nuclei as well as EDFs employed in such
studies~\cite{Rutz1998,Rutz1999,Cwiok1999,Bender2000a,Duguet2001a,Duguet2002a,Bonneau2007a,%
Perez2008,PEREZMARTIN2009,Bertsch2009,Afanasjev2010,Schunck2010,Rodriguez-Guzman2010,%
Rodriguez-Guzman2010a,Dobaczewski2015,Rodriguez-Guzman2016a,Rodriguez-Guzman2017,Ryssens2022}.

The Barcelona Catania Paris Madrid (BCPM) energy density functional 
\cite{Baldo2013} represents a fruitful alternative to more traditional 
phenomenological approaches like the Skyrme or Gogny family of forces. It 
delivers low energy nuclear structure results for even-even nuclei of 
the same quality as  those obtained with the well-known Gogny D1M. The 
relatively low computational effort required by mean field calculations 
with BCPM has fostered a number of applications of BCPM, including
large-scale calculations of finite nuclei
properties~\cite{Giuliani2013,Giuliani2018} and applications to
$r$-process nucleosynthesis calculations~\cite{Giuliani2020} and neutron star
physics~\cite{Sharma2015}
(see~\cite{Baldo2023} for a recent review on BCPM). However, 
time-odd fields are not considered in the functional, preventing the 
description of odd-mass systems as well as single particle excitations 
using the traditional blocking method of the Hartree-Fock-Bogoliubov 
(HFB) method. BCPM is 
built on polynomial fits to realistic nuclear matter equations of state 
for symmetric and pure neutron systems given as a function of the 
density. The same polynomial form is used to define an energy density 
functional for finite nuclei just by replacing the nuclear matter 
density with the density of finite nuclei. A Gaussian potential is 
introduced in the direct channel to simulate surface effects and 
finally, the traditional zero-range spin-orbit plus Coulomb potential 
are added to define the particle-hole component of the functional. A 
density-dependent pairing term~\cite{Garrido1999} completes the 
functional.  In the original formulation of BCPM, polarization effects  at the 
nuclear matter level were not considered and, therefore, the functional 
is only valid for time-reversal symmetry-preserving physical cases, 
such as the ground state of even-even nuclei. This limitation prevents the 
use of BCPM to describe properties of odd-$A$ and odd-odd nuclei as 
well as multi-quasiparticle excitations by means of the traditional
HFB blocking method. On the other hand, by 
comparing different  theoretical approaches describing odd-$A$
nuclei~\cite{Schunck2010}, it has been argued that the time-reversal symmetry 
preserving equal filling approximation (EFA)~\cite{Perez2008} is a valid 
approximation to full quasiparticle blocking. This empirical conclusion 
is based on calculations using zero-range Skyrme forces and therefore 
its range of validity is somehow limited. The main effect of blocking 
is to quench pairing correlations, reducing thereby the pairing gap and 
the typical excitation energy of one-quasiparticle states. The EFA 
leads to the same quenching of pairing correlations even at the 
quantitative level. On the other hand, the EFA can be viewed as a full 
blocking calculation in which time-odd fields in the Hartree-Fock and 
pairing fields generated by the unpaired nucleon
are neglected~\cite{Schunck2010}. The good 
performance of the EFA in describing odd-$A$ nuclei with Skyrme forces, 
and the absence of time-odd fields in its formulation, suggests that EFA 
could be used to extend the realm of the BCPM functional to describe 
odd-$A$ nuclei and quasiparticle excitations. The motivation for this paper 
is threefold: First, we show how the approximate equivalence between EFA and full blocking is also valid with 
Gogny forces, in order to reinforce the empirical conclusion extracted from
Skyrme calculations~\cite{Schunck2010}. Second, we show how to implement 
multi-quasiparticle excitation in the standard variational EFA framework. 
Finally, we assess the accuracy of BCPM-EFA predictions by comparing against
both Gogny
D1M predictions and experimental data of spin 
and parity of ground and low-lying excited states, in a set of selected nuclei 
scattered across the nuclear chart. Multiquasiparticle excitations are also
considered in the paradigmatic case of high-K isomeric states in $^{178}$Hf.

The paper is organized as follows. The theoretical EFA framework and 
the BCPM interaction are outlined in Section~\ref{sec:method}. 
Section~\ref{sec:EFAvsFB} is devoted to a comparison between full 
blocking and EFA calculations. In Section~\ref{sec:results}, the EFA 
results of low-energy spectra and high-$K$ states are presented. 
Conclusions and perspectives for future studies are summarized in 
Section~\ref{sec:summary}.


\section{Theoretical methodology \label{sec:method}}


The self-consistent mean field description of odd-$A$ nuclei, as well as 
that of multi-quasiparticle excitation, is based on the HFB method with 
blocking. In the standard HFB method 
\cite{RS80}, the concept of quasiparticle is introduced by means of the 
Bogoliubov transformation 
\begin{equation} \label{eq:Bogo} 
\left(\begin{array}{c} \beta\\ \beta^{\dagger} 
\end{array}\right)=\left(\begin{array}{cc} U^{\dagger} & V^{\dagger}\\ V^{T} & 
U^{T} \end{array}\right)\left(\begin{array}{c} c\\ c^{\dagger} 
\end{array}\right)\equiv W^{+ }\left(\begin{array}{c} c\\ c^{\dagger} 
\end{array}\right). 
\end{equation} 
The associated HFB state 
$|\Phi\rangle$, is defined as the vacuum state for all the annihilation 
quasiparticle operators $\beta_{\mu}$, i.e.,  
$\beta_{\mu}|\Phi\rangle=0$ for labels $\mu$ indexing all the 
quasiparticle configurations. An important concept in the HFB theory is 
``number parity'', which is nothing but the parity (even or odd) of the 
number of particles in $|\Phi\rangle$ and its excitations. Number 
parity is a symmetry of the system, and wave functions with opposite 
values cannot be mixed together. An even-even nucleus has to be 
described by a HFB state with even number parity for both protons and 
neutrons. The quasiparticle operators carry ``number parity'' 1 and 
therefore $\beta^{\dagger}_{\mu}|\Phi\rangle$ has opposite ``number parity'' 
to that of $|\Phi\rangle$. Therefore, genuine excitations of a given 
system $|\Phi\rangle$ are given by two, four, etc. quasiparticle 
excitations whereas one, three, etc. quasiparticle excitations 
correspond to a neighbor odd-mass system if $|\Phi\rangle$ contains an 
even number of particles (and vice versa). As a consequence, odd-$A$ nuclei have to be treated with a 
``blocked'' wave function $\beta^{\dagger}_{\mu_{B}}|\Phi\rangle$ breaking 
time reversal invariance because of its dependence on the quantum number 
$\mu_{B}$.

For fully paired configurations $|\Phi\rangle$ (i.e., no blocking), 
Bogoliubov amplitudes $U$ and $V$ are determined by using the 
variational principle on the HFB energy $E_\textup{HFB} = \langle \Phi | 
\hat{H} |\Phi \rangle$ leading to the well know HFB equation (see 
\cite{RS80} for a gentle description of the HFB method). Conversely, the  
$U$ and $V$ for a blocked state are obtained by minimizing 
$E_\textup{HFB}^{(\mu_{B})} = \langle \Phi |\beta_{\mu_{B}} \hat{H} 
\beta^{\dagger}_{\mu_{B}}|\Phi \rangle$. The blocking procedure introduces an 
additional level of complexity in the minimization process due to the 
dependence of the energy on $\mu_{B}$. One must consider the 
variational principle for $U$ and $V$ considering all possible values 
of the quantum numbers $\mu_{B}$ of the blocked levels and choosing for 
the ground state the value of $\mu_{B}$ leading to the lowest energy. 
Taking into account that $E_\textup{HFB}^{(\mu_{B})} = \langle \Phi | \hat{H} 
|\Phi \rangle + E_{\mu_{B}} + \cdots $,  where  
$E_{\mu_{B}}$ is the quasiparticle energy of level $\mu_{B}$ and the ellipses 
represent residual interaction corrections, one can expect that a 
search limited to those blocking quasiparticles $\beta^{\dagger}_{\mu_{B}}$ 
with small values of the quasiparticle energies $E_{\mu}$ will end up 
in the lowest energy solution. Typically, considering the ten lowest 
quasiparticle energies leads with confidence to the ground state of the 
odd-$A$ system. On the other hand, the advantage of the above procedure 
is that one obtains not only the ground state but also the lowest energy 
single particle spectrum. 

Wick's theorem also holds for blocked HFB states, and therefore the 
blocked energy $E_\textup{HFB}^{(\mu_{B})} = \langle \Phi 
|\beta_{\mu_{B}} \hat{H} \beta^{\dagger}_{\mu_{B}}|\Phi \rangle$ can be 
expressed in the usual form in terms of the blocked density matrix 
$\rho^{{\mu_{B}}}$ and pairing tensor $\kappa^{{\mu_{B}}}$ given by 
\begin{equation}
\begin{split} \label{eq:roB}
\rho_{kk'}^{(\mu_{B})} & =
\langle\Phi|\beta_{\mu_{B}}c_{k'}^{\dagger}c_{k}
\beta_{\mu_{B}}^{\dagger}|\Phi\rangle \\ 
& =\left(V^{*}V^{T}\right)_{kk'}+ \left(U_{k'\mu_{B}}^{*}U_{k\mu_{B}}-V_{k'\mu_{B}}V_{k\mu_{B}}^{*}\right)
\end{split}
\end{equation}
and
\begin{equation}
\begin{split} \label{eq:KaB}
\kappa_{kk'}^{(\mu_{B})} & =
\langle\Phi|\beta_{\mu_{B}}c_{k'}c_{k}\beta_{\mu_{B}}^{\dagger}|\Phi\rangle \\ 
& = \left(V^{*}U^{T}\right)_{kk'}+
\left(U_{k\mu_{B}}V_{k'\mu_{B}}^{*}-
U_{k'\mu_{B}}V_{k\mu_{B}}^{*}\right).
\end{split}
\end{equation}
Owing to their dependence on the quantum number $\mu_{B}$, these two matrices 
are not time-reversal invariant. In order to restore time-reversal symmetry, the EFA~\cite{Perez2008}
postulates the use of the ``averaged'' density
\begin{equation}
\begin{split}
\rho_{kk'}^\textup{EFA}  =
\left(V^{*}V^{T}\right)_{kk'}  +
\frac{1}{2} \Bigl( &U_{k'\mu_{B}}U_{k\mu_{B}}^{*}-
	V_{k'\mu_{B}}^{*}V_{k\mu_{B}} \Bigr. \\ 
\Bigl. + &U_{k'\overline{\mu}_{B}}
U_{k\overline{\mu}_{B}}^{*}-
V_{k'\overline{\mu}_{B}}^{*}V_{k\overline{\mu}_{B}}\Bigr) \,,
\label{eq:ROEFA}
\end{split}
\end{equation}
and ``average'' pairing tensor 
\begin{equation}
\begin{split}
\kappa_{kk'}^\textup{EFA} =\left(V^{*}U^{T}\right)_{kk'} 
  + \frac{1}{2}
  \Bigl(& U_{k\mu_{B}}V_{k'\mu_{B}}^{*}-U_{k'\mu_{B}}V_{k\mu_{B}}^{*} \Bigr. \\ 
+ \Bigl. 
& U_{k\overline{\mu}_{B}}V_{k'\overline{\mu}_{B}}^{*}-
U_{k'\overline{\mu}_{B}}V_{k\overline{\mu}_{B}}^{*}\Bigr)\label{eq:KAEFA} \,,
\end{split}
\end{equation}
instead. The EFA energy is obtained by replacing 
$\rho_{kk'}^{(\mu_{B})}$ and $ \kappa_{kk'}^{(\mu_{B})}$ with their EFA 
equivalents in the expression of the blocked energy 
$E_\textup{HFB}^{(\mu_{B})}$. It was shown in Ref~\cite{Perez2008} that 
this procedure can be justified by using quantum statistical mechanic 
concepts. The EFA corresponds to a statistical admixture of the states 
$ \beta^{\dagger}_{\mu_{B}}|\Phi \rangle$ and $ 
\beta^{\dagger}_{\bar{\mu}_{B}} |\Phi \rangle$ with equal probability 
1/2. Traditionally, quantum statistical admixtures are better described 
by using a density matrix operator given by the statistical ensemble 
considered. However, arbitrary statistical distributions can also be 
handled by introducing a density matrix operator $\hat{\mathcal{D}}$ 
chosen in such a way that $\hat{\mathcal{D}}|\Phi\rangle=|\Phi\rangle$ 
and 
$\hat{\mathcal{D}}\beta_{\mu}^{\dagger}=p_{\mu}\beta_{\mu}^{\dagger}\hat{\mathcal{D}}$, 
$p_{\mu}$ being the probability of the one-quasiparticle excitation 
$\beta_{\mu}^{\dagger}|\Phi\rangle$~\cite{Perez2008}. In this 
formalism, the statistical mean value of an arbitrary  operator is 
given by 
\begin{equation}
\begin{split}
\langle\hat{O}\rangle_{S} & =
\frac{\textrm{Tr}[\hat{O}\hat{\mathcal{D}}]}{\textrm{Tr}[\hat{\mathcal{D}}]} 
 = \frac{1}{Z}\left(\left\langle \Phi\right|\hat{O}\left|\Phi\right\rangle +
   \sum_{\mu}p_{\mu}\left\langle \Phi\right|\beta_{\mu}\hat{O}\beta_{\mu}^{\dagger}\left|\Phi\right\rangle\right. \\ 
 & + \left. \frac{1}{2!}\sum_{\nu\mu}p_{\mu}p_{\nu}\left\langle \Phi\right|\beta_{\mu}\beta_{\nu}\hat{O}\beta_{\nu}^{\dagger}\beta_{\mu}^{\dagger}\left|\Phi\right\rangle \ldots\right)
 \label{eq:MVS}
\end{split}
\end{equation}
with
\begin{equation}
\textrm{Tr}[\hat{\mathcal{D}}]=Z=1+\sum_{\mu}p_{\mu}+\sum_{\nu<\mu}p_{\mu}p_{\nu}\ldots=\prod_{\mu}(1+p_{\mu})
\label{eq:Z}
\end{equation}
It is worthwhile to remark that in the statistical framework there are contributions
from terms with different ``number parity''.  However, it has been
shown by using ``number parity'' projection techniques that the contamination
is not relevant for describing the physics~\cite{Delft1996,Rossignoli1998,Esashika2005}.
Thanks to the existence of a statistical Wick's theorem (see for instance
the proof given by ~\citet{Gaudin1960} or~\citet{PerezMartin2007} for
a more recent account), it is possible to compute any statistical mean
value of a product of creation and annihilation operators in terms
of the corresponding contractions. Therefore, it is possible to
express the statistical mean value of the energy 
$\langle\hat{H}\rangle_{S}=\textrm{Tr}[\hat{H}\hat{\mathcal{D}}]/\textrm{Tr}[\hat{\mathcal{D}}]$
by using the standard expression 
\begin{equation}
\langle\hat{H}\rangle_{S}=\mathrm{Tr}[t\rho]+\frac{1}{2}\mathrm{Tr}[\Gamma\rho]-\frac{1}{2}\mathrm{Tr}[\Delta\kappa^{*}]
\,,
\label{eq:HFBE}
\end{equation}
where $\Gamma$ is the mean-field potential, $\Delta$ the pairing field, 
and the density matrix and pairing tensor are given by the contractions
\begin{equation}
\rho_{kk'}=\frac{\textrm{Tr}(c_{k'}^{\dagger}c_{k}\hat{\mathcal{D}})}{\textrm{Tr}(\hat{\mathcal{D}})};\:\:\kappa_{kk'}=\frac{\textrm{Tr}(c_{k'}c_{k}\hat{\mathcal{D}})}{\textrm{Tr}(\hat{\mathcal{D}})}
\,.
\label{eq:Cont}
\end{equation}
In the EFA case, the probabilities are given by 
\begin{equation}
p_{\mu}=\left\{ \begin{array}{ccc}
		1 &  & \text{if $\mu=\mu_{B}$ or $\overline{\mu}_{B}\,$;}\\
0 &  & \text{otherwise.}\end{array}\right.\label{eq:P_EFAprob}
\end{equation}
Inserting these probabilities into the contractions of Eq.~\eqref{eq:Cont}
and taking into account the definition of the statistical average Eq.~\eqref{eq:MVS}, 
one comes to the conclusion that the probabilities of Eq.~\eqref{eq:P_EFAprob}
lead to the density matrix and pairing tensors of the EFA Eqs.~\eqref{eq:ROEFA} and \eqref{eq:KAEFA}. As a 
consequence, the statistical average of the energy with the probabilities of 
Eq.~(\ref{eq:P_EFAprob}) is nothing but the EFA energy 
\begin{equation}
	E_\textup{EFA}=\textrm{Tr}[\hat{H}\hat{\mathcal{D}}^\textup{EFA}]/\textrm{Tr}[\hat{\mathcal{D}}^\textup{EFA}]\,.
\end{equation}
This result justifies the otherwise \emph{ad hoc} expression of the EFA 
energy and provides its interpretation as the statistical mean value of 
the Hamiltonian taken with the EFA density operator. By using Eq.~(\ref{eq:MVS}) 
together with Eq.~(\ref{eq:P_EFAprob}), the EFA energy 
$E_\textup{EFA}$ can also be written in a more transparent way  as
\begin{equation}
\begin{split}
E_\textup{EFA} & =\frac{1}{4}\left(\left\langle \Phi\right|\hat{H}\left|\Phi\right\rangle +\left\langle \Phi\right|\beta_{\mu_{B}}\hat{H}\beta_{\mu_{B}}^{\dagger}\left|\Phi\right\rangle \right. \\ 
& +\left. \left\langle \Phi\right|\beta_{\overline{\mu}_{B}}\hat{H}\beta_{\overline{\mu}_{B}}^{\dagger}\left|\Phi\right\rangle +\left\langle \Phi\right|\beta_{\mu_{B}}\beta_{\overline{\mu}_{B}}\hat{H}\beta_{\overline{\mu}_{B}}^{\dagger}\beta_{\mu_{B}}^{\dagger}\left|\Phi\right\rangle \right) \,.
\label{eq:E_EFAa}
\end{split}
\end{equation}
This expression shows that the EFA energy is simply an average with
equal weights of the energy of the reference even-even wave function
$|\Phi\rangle$, the energies of one quasi-particle excitations with
quantum numbers $\mu_{B}$ and $\overline{\mu}_{B}$, and the energy
of the two quasi-particle excitation with the same quantum numbers.
This result is very illustrative of the nature of the EFA as a statistical
theory. The same kind of arguments can be applied to compute mean
values of any kind of operators in the EFA framework. An important result
that can be easily derived is that the EFA mean values of any one-body
operator, which according to the general result can be written as
\begin{equation}
\begin{split}
\langle\hat{O}\rangle_\textup{EFA} & =\frac{1}{4}\left(\left\langle \Phi\right|\hat{O}\left|\Phi\right\rangle +\left\langle \Phi\right|\alpha_{\mu_{B}}\hat{O}\alpha_{\mu_{B}}^{\dagger}\left|\Phi\right\rangle \right.  \\ 
& +\left. \left\langle \Phi\right|\alpha_{\overline{\mu}_{B}}\hat{O}\alpha_{\overline{\mu}_{B}}^{\dagger}\left|\Phi\right\rangle +\left\langle \Phi\right|\alpha_{\mu_{B}}\alpha_{\overline{\mu}_{B}}\hat{O}\alpha_{\overline{\mu}_{B}}^{\dagger}\alpha_{\mu_{B}}^{\dagger}\left|\Phi\right\rangle \right),
\label{eq:MVQP}
\end{split}
\end{equation}
can also be written in a more compact form as
\begin{equation}
\begin{split}
\langle\hat{O}\rangle_\textup{EFA} & =\frac{1}{2}\left(\left\langle \Phi\right|\alpha_{\mu_{B}}\hat{O}\alpha_{\mu_{B}}^{\dagger}\left|\Phi\right\rangle +\left\langle \Phi\right|\alpha_{\overline{\mu}_{B}}\hat{O}\alpha_{\overline{\mu}_{B}}^{\dagger}\left|\Phi\right\rangle \right) \\ 
& =\frac{1}{2}\left(\left\langle \Phi\right|\hat{O}\left|\Phi\right\rangle +\left\langle \Phi\right|\alpha_{\mu_{B}}\alpha_{\overline{\mu}_{B}}\hat{O}\alpha_{\overline{\mu}_{B}}^{\dagger}\alpha_{\mu_{B}}^{\dagger}\left|\Phi\right\rangle \right) \,.
\label{eq:MVQPC}
\end{split}
\end{equation}
This allows us to write the density matrix and pairing tensor as an
average over one quasi-particle excitations
\begin{equation}
\rho_{kk'}^\textup{EFA}=\frac{1}{2}\left(\langle\Phi|\alpha_{\mu_{B}}c_{k'}^{\dagger}c_{k}\alpha_{\mu_{B}}^{\dagger}|\Phi\rangle+\langle\Phi|\alpha_{\bar{\mu}_{B}}c_{k'}^{\dagger}c_{k}\alpha_{\bar{\mu}_{B}}^{\dagger}|\Phi\rangle\right) \,,
\label{eq:roEFAC}
\end{equation}
and 
\begin{equation}
\kappa_{kk'}^\textup{EFA}=\frac{1}{2}\left(\langle\Phi|\alpha_{\mu_{B}}c_{k'}c_{k}\alpha_{\mu_{B}}^{\dagger}|\Phi\rangle+\langle\Phi|\alpha_{\bar{\mu}_{B}}c_{k'}c_{k}\alpha_{\bar{\mu}_{B}}^{\dagger}|\Phi\rangle\right)\,,
\label{eq:kaEFAC}
\end{equation}
which is a very intuitive result according to the expressions of Eqs.
\eqref{eq:ROEFA} and \eqref{eq:KAEFA}. This result, however, does
by no means imply that the energy, which is the average of a two-body
operator, could be written as 
$\frac{1}{2}\left(\langle\Phi|\alpha_{\mu_{B}}H\alpha_{\mu_{B}}^{\dagger}|\Phi\rangle+\langle\Phi|\alpha_{\bar{\mu}_{B}}H\alpha_{\bar{\mu}_{B}}^{\dagger}|\Phi\rangle\right)$.

Once the EFA energy is defined in terms of the density matrix and the pairing
tensor, and those objects in terms of the Bogoliubov $U$ and $V$ amplitudes
of Eq.~\eqref{eq:Bogo}, the application of the variational principle is 
straightforward, and it has been discussed at length in~\cite{Perez2008}. 
As in the case of other density functionals (Gogny, Skyrme, etc.) the 
rearrangement terms coming from the derivatives of the density are taken
into account in the HFB equation.
This is also the case for the BCPM calculations for even-even nuclei. 

For higher order excitations including two, three, etc. quasiparticle excitations,
one can proceed along the same lines as discussed above. For multi-quasiparticle 
excitations the density matrix of Eq.~\eqref{eq:roB} becomes
\begin{equation}
\begin{split}\label{eq:roBmqp}
\rho_{kk'}^{(\mu_{B_{1}},\ldots,\mu_{B_{N}})} & =
\langle\Phi| \left( \prod_{\sigma}\beta_{\sigma} \right) c_{k'}^{\dagger}c_{k}
\left( \prod_{\sigma}\beta_{\sigma}^{\dagger} \right) |\Phi\rangle \\ 
& =\left(V^{*}V^{T}\right)_{kk'} + \sum_{\sigma} \left(U_{k'\sigma}^{*}U_{k\sigma}-V_{k'\sigma}V_{k\sigma}^{*}\right) \,,
\end{split}
\end{equation}
where the index $\sigma$ runs over the set $\mu_{B_{1}},\ldots,\mu_{B_{N}}$ for a
$N$-quasiparticle excitation. With respect to the one quasiparticle case,
it amounts to replacing the term in parenthesis on the right hand side
of Eq.~\eqref{eq:roB} by the sum on the right hand side of Eq.~\eqref{eq:roBmqp}.
The corresponding expression for the $\kappa$ pairing tensor can be easily
obtained from Eq.~\eqref{eq:KaB}.

The EFA expression for the multi-quasiparticle excitation density is obtained
from Eq.~\eqref{eq:roBmqp} by multiplying the sum on the right hand side by
one half and extending the sum on the label $\sigma$ to include the
time reverse quantum numbers of the set $\mu_{B_{1}},\ldots,\mu_{B_{N}}$. 
The same consideration applies straightforwardly to the EFA pairing tensor. 
The EFA density matrix and pairing tensors can be obtained from Eq.~(\ref{eq:Cont}) by introducing
the statistical probabilities
\begin{equation}
p_{\sigma}=\left\{ \begin{array}{ccc}
1 &  & \text{$\sigma \in {\mu_{B_{1}},\ldots,\mu_{B_{N}}}$; or $\in {\overline{\mu}_{B_{1}},\ldots,\overline{\mu}_{B_{N}}}$;} \\
0 &  & \text{otherwise.}\end{array}\right.\label{eq:P_EFAprobMQP}
\end{equation}
For instance, in the two-quasiparticle case, and according to Eq.~(\ref{eq:MVQP}) 
the EFA energy is the sum of multiple terms, most of them differing by mere
permutations of the labels. Once this multiplicity is taken into account,
one obtains
\begin{equation}
\begin{split}\label{eq:E_EFAa2qp}
E_\textup{EFA}  =\frac{1}{16} \Big( & \bullet + \mu \bullet \mu + \nu \bullet \nu + \overline{\mu} \bullet \overline{\mu} + \overline{\nu} \bullet \overline{\nu} \\  
& + \mu \nu \bullet \nu \mu  + \mu \overline{\mu} \bullet  \overline{\mu} \mu + \nu \overline{\nu} \bullet \overline{\nu} \nu \\  
& + \mu \overline{\nu} \bullet \overline{\nu} \mu + \nu \overline{\mu} \bullet \overline{\mu} \nu + \overline{\mu} \overline{\nu} \bullet \overline{\nu} \overline{\mu} \\  
& + \mu \overline{\mu} \nu \bullet \nu \overline{\mu} \mu +  \mu \overline{\mu} \overline{\nu} \bullet \overline{\nu} \overline{\mu} \mu \\ 
&  + \mu \nu \overline{\nu} \bullet \overline{\nu} \nu \mu +  \overline{\mu} \nu \overline{\nu} \bullet \overline{\nu} \nu \overline{\mu} \\  
& + \mu \overline{\mu} \nu \overline{\nu} \bullet \overline{\nu} \nu \overline{\mu} \mu   \Big) \,,
\end{split}
\end{equation}
where, in order to simplify the expression, we have introduced an obvious
notation. For instance, the term $\nu \bullet \nu$ represents the overlap $ \langle \Phi | \beta_{\nu_{B}}\hat{H}\beta_{\nu_{B}}^{\dagger} | \Phi \rangle  $.
The average contains 16 terms. 
Moreover, the average is manifestly invariant under time-reversal. As in the one-quasiparticle
case, the average contains contributions from even and odd ``number parity''
states. In order to understand the contents of the previous average, it is
convenient to use the quasiparticle representation of the Hamiltonian (see Appendix E of~\cite{RS80})
with the quasiparticle operators defined in such a way as to render the
$H^{11}=\sum_{\sigma} E_{\sigma} \beta^{\dagger}_{\sigma} \beta_{\sigma}$ term diagonal. Using this 
representation, Eq.~\eqref{eq:E_EFAa2qp} becomes
\begin{equation}
	E_\textup{EFA} = \langle \Phi|\hat{H}|\Phi\rangle + \frac{1}{2} \big(
	E_{\mu_{B}} + E_{\overline{\mu}_{B}}+ E_{\nu_{B}}+E_{\overline{\nu}_{B}} \big) +
	\cdots
\end{equation}
where the missing terms are contractions of the $H^{22}$ part of the Hamiltonian.

\subsection{The BCPM functional}

The BCPM functional was proposed in Ref~\cite{Baldo2013} as an 
evolution of the BCP functional~\cite{Baldo2008,Baldo2010} to include 
several improvements aimed to reduce the original number of adjustable 
parameters. The idea was to use state-of-the-art microscopic nuclear 
matter calculations with realistic nuclear forces to obtain a realistic 
equation of state as a function of the density. To convert the 
numerical tables into some analytical form, a polynomial fit up to fifth 
order was used to fit the equation of state up to six times saturation 
density~\cite{Baldo2008}. The same two polynomials (one for pure nuclear matter, the 
other for neutron matter) are used in finite nuclei replacing nuclear 
matter densities by the densities of finite nuclei. The functional is 
supplemented with a Gaussian term in the direct channel to simulate 
surface effects, a spin-orbit term with the same functional form as the 
Skyrme or Gogny interactions, the Coulomb potential among protons, and 
a density-dependent pairing term; see the recent review~\cite{Baldo2023} for a 
more detailed explanation of the properties of BCPM and its application 
to nuclear structure problems~\cite{Giuliani2013,Giuliani2014,Neufcourt2019a,Neufcourt2020,Giuliani2018} and
astrophysics~\cite{Sharma2015,Giuliani2020}. As mentioned in the introduction, the 
main deficiency of BCPM is the lack of time-odd densities, preventing 
its use in those situations requiring the breaking of time-reversal 
invariance. Therefore, high spin physics, odd-mass systems, or 
multiquasiparticle excitations cannot be treated in the framework of 
BCPM\@. However, the fact that the EFA and full blocking provide similar 
results in calculations with Skyrme interactions suggests that the EFA 
can be used along with BCPM to define a functional for odd-mass systems or 
quasiparticle excitations. Based on the results with Skyrme and a 
comparison with Gogny forces described below, one can expect that the 
differences between the results obtained in the present approach and 
those obtained in a full blown blocking procedure are not going to be 
relevant in terms of the physics to be described.


\subsection{Orthogonality}

The variational character of the EFA is advantageous in many instances, 
but it also presents disadvantages in other aspects of the calculation. 
It is usually found in real calculations that most of the blocked 
configurations end up converging to the lowest energy configuration 
with the same set of quantum numbers. The only situation when this is 
no longer the case is when the initial blocked configuration has 
deformation parameters very different from the ones of the lowest 
energy solution, and therefore the iterative process starts in the 
neighborhood of a local minimum that cannot be avoided by the gradient 
method. This unwanted property implies that in a calculation breaking all
symmetries it is very likely that only one blocked configuration is going 
to be reached, irrespective of the initial blocking configuration. This 
drawback of the method can be circumvented by introducing orthogonality 
constraints in the calculation~\cite{Egido1980}. However, it is easier and usually 
provides more physics insight to preserve axial symmetry 
in order to have $K$ (the component of angular momentum along the 
third axis) as a good quantum number. In this way, orthogonality among 
different $K$ values is automatic and therefore one is guaranteed at least 
one solution for each possible value of $K$. Additionally, if octupole 
deformation is not relevant, it is very likely that parity is also going 
to be preserved, giving two states for each $K$ value. 

In the results discussed below, axial symmetry is preserved and $K$ is
a good quantum number. Reflection symmetry is allowed to break but in
many instances, the solution preserves this symmetry to a large extent and
therefore parity can be used as an ``approximate'' quantum number (the mean
value of parity is computed and taking as a good quantum number if its
absolute value is above 0.9). On the other hand, the orthogonality constraint
is well defined in full blocking calculations and it has been implemented
in all the examples considered. This is not the case with the EFA as the
concept of orthogonality in a statistical ensemble is not well defined. 

\section{EFA versus full blocking \label{sec:EFAvsFB}}

The equivalence between EFA and full blocking calculations has been 
empirically tested only in Skyrme forces~\cite{Schunck2010}. It is the 
purpose of this section to extend the validity of the equivalence also to Gogny forces 
by presenting results obtained in both approaches with Gogny D1M for 
some selected set of nuclei, the same used in the next section to 
empirically demonstrate the convenience of using EFA to describe 
odd-mass nuclei in BCPM\@. The choice of the Gogny force is not 
accidental and is based on its good pairing properties as well as its good
description of high-spin physics~\cite{Robledo2019}, that critically relies on appropriate 
time-odd component of the force and pairing properties. For the full blocking calculation with the 
Gogny force, we use the results of Refs~\cite{Robledo2012,Robledo2014}. 
For auxiliary HFB calculations for even-even systems we 
follow~\cite{Robledo2011}. For other examples of EFA calculations with 
the Gogny force see 
\cite{Rodriguez-Guzman2010,Rodriguez-Guzman2010a,Rodriguez-Guzman2010b,PEREZMARTIN2009,Rodriguez-Guzman2011,Rodriguez-Guzman2017}.

\begin{figure}[tb]
\begin{center}
\includegraphics[width=\columnwidth]{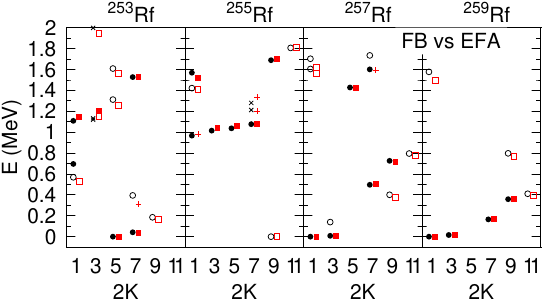}
\caption{The lowest energy excitation spectra for several rutherfordium isotopes
computed with the Gogny D1M force, using full blocking and the EFA method.
Each state is represented by a symbol according to its parity and the method
employed for its calculation. Symbols are placed along the horizontal axis according
to the value of $2K$. Filled (empty) black circles represent full blocking results with
positive (negative) parity. States with octupole deformation and no definite
parity are depicted with a black $\times$ symbol. Filled (empty) red squares are used for
positive (negative) states obtained with 
EFA\@. States with octupole deformation and no definite parity are depicted
with a red $\color{red}+$ symbol.
All symbols are slightly displaced along the horizontal axis to improve
legibility.} 
\label{fig:Rf-FB-EFA}
\end{center}
\end{figure}

In Fig.~\ref{fig:Rf-FB-EFA} a comparison of the results obtained with the Gogny D1M using
full blocking and the EFA  is shown for $^{253-259}$Rf isotopes. 
As can be clearly noticed, the correspondence between the
EFA and full blocking results is one to one and remarkable. These
results go along the conclusions of~\cite{Schunck2010} for Skyrme forces
and confirm the nearly perfect equivalence between EFA and full blocking
in terms of excitation energies. All the levels have prolate quadrupole
deformation with $\beta_{2}\approx 0.28$ and negligible (except in some
specific levels) octupole deformation. Please note that some levels
obtained with full blocking do not have their counterpart in the EFA calculation.
This is because in the EFA the orthogonality constraint has not been imposed. 

\begin{figure}[tb]
\begin{center}
\includegraphics[width=\columnwidth]{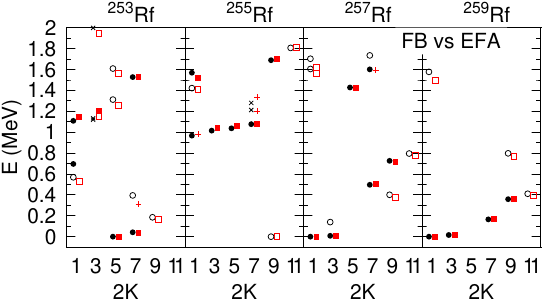}
\caption{A comparison of results obtained for Rf isotopes with 
full blocking and the Gogny force with D1S and D1M parametrizations. For an
interpretation of the plot, see caption of Fig.~\ref{fig:Rf-FB-EFA}. In the
present case, black circles (red squares) represent D1S (D1M) calculations.}
\label{fig:Rf_D1S-D1M}
\end{center}
\end{figure}

It is also interesting to consider the different results obtained with
different parametrizations of the Gogny force. It is well known in the
literature that the excitation energy of different excited states in odd-mass nuclei
is spread over a broad range of values depending on the force/functional used 
\cite{Afanasjev2011,Rodriguez-Guzman2010a,Rodriguez-Guzman2010b,Rodriguez-Guzman2011,Dobaczewski2015}.
In Fig.~\ref{fig:Rf_D1S-D1M} we compare our results with full blocking
for D1S and D1M and the rutherfordium isotopes discussed above. In line with previous
findings, we observe some dispersion in the excitation energies that is 
most likely caused by slight changes in the position of the underlying
single particle levels or slightly different pairing correlations. It is
clear that reaching the so called ``spectroscopic accuracy'' (in the
present context meaning perfect matching of spectra using different interactions) with forces
containing 14 parameters, most of them adjusted to nuclear matter properties
and aimed to be valid across the nuclear chart, is unfeasible. 
Spectroscopic accuracy can only be achieved by means of interactions fitted
to very specific regions of the nuclear chart, as is customary in
shell model calculations. 

\begin{figure}[tb]
\begin{center}
\includegraphics[width=\columnwidth]{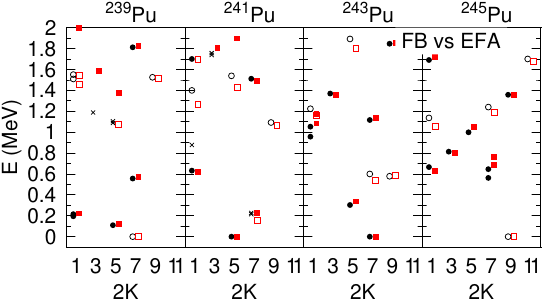}

\vspace{0.2cm}

\includegraphics[width=\columnwidth]{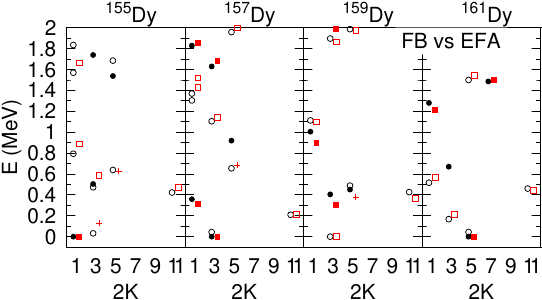}

\vspace{0.2cm}

\includegraphics[width=\columnwidth]{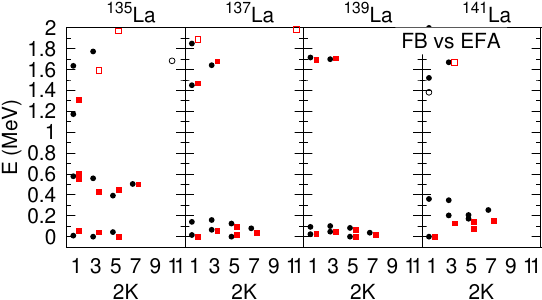}
\caption{A comparison of full blocking and EFA results for selected 
Pu, Dy and La isotopes. For an
interpretation of the plot, see caption of Fig.~\ref{fig:Rf-FB-EFA}.} 
\label{fig:Pu-FB-EFA}
\end{center}
\end{figure}

In Fig.~\ref{fig:Pu-FB-EFA} results obtained with the Gogny D1M using 
full blocking are compared to the EFA ones for plutonium, dysprosium and lanthanum 
isotopes. As in the rutherfordium case, the agreement is very good. Plutonium
isotopes have a quadrupole deformation parameter 
$\beta_{2}\approx 0.27$ in both the full blocking and EFA cases. Some 
of the levels obtained with full blocking have octupole deformation 
with $\beta_{3}$ as large as 0.1 in $^{239}$Pu. In dysprosium isotopes, the 
deformation goes from $\beta_{2}\approx0.20$ in $^{155}$Dy up to 
$\beta_{2}\approx0.30$ in $^{161}$Dy. Octupole deformation is 
negligible in all the instances, being slightly larger in the EFA\@. In 
the odd-proton set of lanthanum isotopes, there is shape coexistence in 
$^{135}$La with minima located at $\beta_{2}\approx0.10$ in the prolate 
side, and $\beta_{2}\approx -0.10$ in the oblate one.  The two isotopes 
$^{137-139}$La are spherical, whereas $\beta_{2}\approx 0.06$ in 
$^{141}$La. Again, octupole deformation is negligible. In the EFA case, 
$^{135}$La and $^{141}$La are slightly prolate with 
$\beta_{2}\approx0.12$ and $\beta_{2}\approx0.07$, respectively. As in 
the full blocking case,  $^{137-139}$La are spherical.

\section{BCPM EFA\label{sec:results}}

When the BCPM functional was postulated~\cite{Baldo2013}, similitude was noted
between the results of Gogny D1M and those of BCPM for low energy nuclear structure
quantities~\cite{Giuliani2013}. Therefore, it seems reasonable to compare the BCPM-EFA
predictions for odd-$A$ nuclei with Gogny D1M ones. The choice of D1M
is further justified by its good performance in
describing time-odd-physics like rotational bands or odd-mass nuclei. 
In the following, we consider five different isotopic chains covering different
regions of the nuclear chart: the superheavy element rutherfordium, the actinide
plutonium, the rare-earth elements lanthanum and dysprosium, and the light
krypton isotopes. Overall, these five isotopic chains cover a large variety of
nuclear deformations and mass numbers, providing a complete benchmark for BCPM-EFA calculations. 

Regarding the comparison of the obtained spectra with experimental data, it has 
to be taken into account that there are many missing effects that could 
influence, to the level of a few hundred keV, the excitation energies 
obtained by the mean field, particle-vibration coupling being a common 
suspect~\cite{Afanasjev2011}. Other factors that could influence the 
position of the levels are the correlation energies associated to 
various symmetry restorations~\cite{Sheikh2021}, like particle number or 
angular momentum and parity, that depend on the 
detailed deformation and pairing properties of the levels.
It is to be expected that the large rotational energy correction (of 
the order of a few~MeV in the present case) can be slightly different 
for different orbitals (pairing correlations are different for each 
orbital and, accordingly, moments of inertia would also be different). 
The same argument applies to the zero-point energy correction 
associated with particle number restoration. Last but not least, 
correlation energies associated to fluctuations in the collective 
degrees of freedom (closely related to the particle-vibration coupling) are also
expected to differ from one orbital to another. A substantial reshuffling of
orbitals may occur after all these corrections have been implemented.  Those
unaccounted effects can modify the ordering of level with respect to the plain
mean field results. Even though this is an exciting field of research and work
along these lines is already in progress, it goes beyond the purpose of this
work. We follow therefore a more pragmatic approach in the
comparison with experimental data, and consider that calculations properly reproduce
the experimental data if the predicted level lies within a window centered at the experimental value. In
this work, we take an empirical width of 300~keV for such energy
window. 

\subsection{Rutherfordium isotopes}   

\begin{figure}[tb]
\begin{center}
\includegraphics[width=\columnwidth]{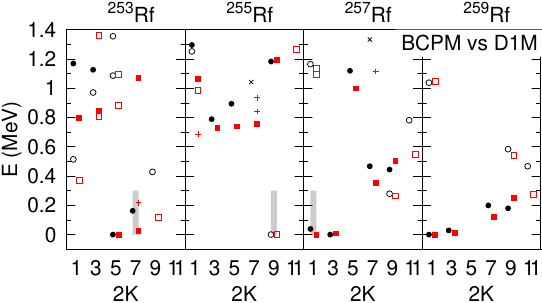}
\caption{The lowest energy spectrum obtained for $^{253-259}$Rf with the BCPM
functional and the D1M Gogny force. For an
interpretation of the plot, see caption of Fig.~\ref{fig:Rf-FB-EFA}. In the
present case, black circles (red squares) represent BCPM (D1M) calculations. 
Experimental ground state $J$ values are marked with a gray bar. The height of
the bar represents the energy window for the comparison with theoretical results (see text for details).} 
\label{fig:Rf}
\end{center}
\end{figure}

In Fig.~\ref{fig:Rf}, the lowest-energy spectra obtained with BCPM and D1M 
for $^{253-259}$Rf isotopes are plotted. Please note
that excitation energies of D1M have been multiplied by a factor 0.75 to
compare with those of BCPM due to different effective masses (0.75 for D1M, 1.0 for BCPM). 
By looking at the
spectra one can easily recognize the close similitude between the two calculations. 
There are some differences that can be attributed to specific positions of single particle
levels, but they are minor and usually do not alter the ordering of the different states. 
The $\beta_{2}$ deformation parameters of all the levels shown belong to the interval $0.27-0.29$
for $^{253-257}$Rf, and to the interval  $0.26-0.28$ for $^{259}$Rf. In most cases, the $\beta_{3}$ octupole deformation parameter is zero and therefore
parity is a good quantum number. 

In both cases the excitation energy reduction with respect to the unperturbed 
one quasiparticle energy is a consequence of the quenching of pairing correlations
due to the ``pseudo-blocking'' effect of EFA. A measure of pairing correlation
strength is the particle-particle energy $\textrm{Tr} (\Delta \kappa) / 2$. In the present
case, while the proton $p-p$ energy remains more or less the same as the one of
the neighbor even-even isotopes, it is quenched by a factor of around 2
in the neutron channel. The quenching factor can reach a value of four in
some cases, like the $7/2^{+}$ state in $^{253}$Rf. Also, pairing correlations
in the neutron channel can disappear as in the case of the $9/2^{-}$ in $^{255}$Rf.
This result indicates that binding energies obtained within the EFA framework may
substantially differ from those from perturbative schemes~\cite{Duguet2001a,Bonneau2007a} 
often used in large scale calculations.

Experimental data in this region is scarce~\cite{Kondev2021}, and all 
the spin and parity assignments are tentative. The isotope $^{253}$Rf 
has a tentative $7/2^{+}$ ground state that could correspond to the 
predicted $7/2^{+}$ observed at around 200~keV excitation energy in 
Fig.~\ref{fig:Rf}. $^{255}$Rf has a tentative $9/2^{-}$ experimental 
ground state in agreement with both predictions. For $^{257}$Rf, the 
lowest states are $1/2^{+}$, $5/2^{+}$, and $11/2^{-}$. In our 
calculations those levels appear at an excitation energy of 20, 
1120 and 794~keV, respectively. For the isotope $^{259}$Rf 
there are no experimental data available. In light of these results we conclude that, despite
the absence of beyond mean-field effects discussed above, the agreement with experimental data 
is satisfactory.

\subsection{Plutonium isotopes}
\begin{figure}[tb]
\begin{center}
\includegraphics[width=\columnwidth]{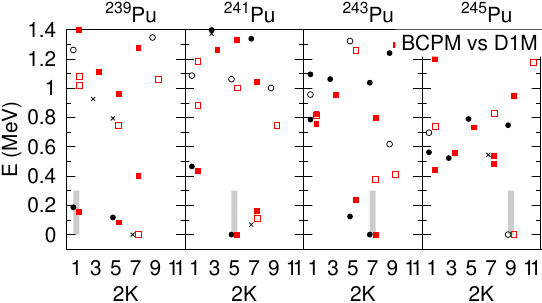}
\caption{The lowest energy spectrum obtained for $^{239-245}$Pu with the BCPM
functional  and the D1M Gogny force. For an
interpretation of the plot, see caption of Fig.~\ref{fig:Rf-FB-EFA}. In the
present case, black circles (red squares) represent BCPM (D1M) calculations.
Experimental ground state $J$ values are marked with a gray bar. The height of
the bar represents the energy window for the comparison with theoretical results (see text for details).}
\label{fig:Pu}
\end{center}
\end{figure}

Results for plutonium isotopes corresponding to mass numbers $A=239-245$ 
are shown in Fig.~\ref{fig:Pu} for both BCPM and Gogny D1M. As in the 
previous case, we find a close similitude in both spectra. The 
experimental ground-state angular momentum and parity taken from Ref 
\cite{Kondev2021} are $J^{\pi}=1/2^{+}$, $J^{\pi}=5/2^{+}$ 
$J^{\pi}=7/2^{+}$, and $J^{\pi}=9/2^{-}$, respectively. Those values are
accurately predicted by both EDFs, with the exception of $J^{\pi}=1/2^{+}$ in
$^{239}$Pu, that appears in the calculations at an excitation energy of around
100~keV (which is within the 300~keV window discussed above).
The ground state has a $\beta_{2}$ deformation of 0.27 in the 
three cases and shows no trace of reflection asymmetry. For the lighter
$^{229-237}$Pu isotopes, all the experimental ground-state
$J^{\pi}$ values~\cite{Kondev2021} are predicted by BCPM within the 300~keV energy window. 
The only exception is for the $J^{\pi}=7/2^{-}$ 
ground state in $^{237}$Pu, which lies at 0.530 MeV excitation energy with BCPM.

\subsection{Dysprosium isotopic chain}

\begin{figure}[tb]
\begin{center}
\includegraphics[width=\columnwidth]{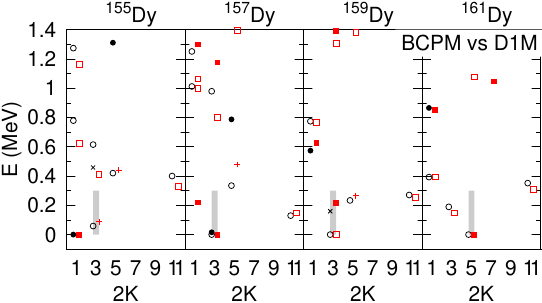}
\caption{The lowest energy spectrum obtained for $^{155-161}$Dy with the BCPM
functional and the D1M Gogny force. For an
interpretation of the plot, see caption of Fig.~\ref{fig:Rf-FB-EFA}. In the
present case, black circles (red squares) represent BCPM (D1M) calculations.
Experimental ground state $J$ values are marked with a gray bar. The height of
the bar represents the energy window for the comparison with theoretical results (see text for details).}
\label{fig:Dy}
\end{center}
\end{figure}

Results for the $^{155-161}$Dy isotopes are displayed in Fig.~\ref{fig:Dy}. 
The four isotopes are prolate, with $\beta_{2}$ ranging from  
0.23 in $^{155}$Dy to  0.32 in $^{161}$Dy. Each of the levels 
displayed in Fig.~\ref{fig:Dy} has its own $\beta_{2}$ deformation, 
that differs by $\pm0.04$ from the quadrupole deformation of 
the ground state.  As in the previous cases, the agreement between 
BCPM and D1M results is remarkable. In the four isotopes  $^{155-161}$Dy there is a 
known isomeric state~\cite{Kondev2021} with $J^{\pi}=11/2^{-}$ at excitation
energies of 234, 199, 352, and 485~keV, respectively. This 
isomeric state is well reproduced by the calculations. In $^{157}$Dy
there is an additional isomeric state,  $J^{\pi}=9/2^{+}$ at 162~keV,
that is not observed in our calculations. Spin and parity of the
ground state is known experimentally for the isotopes $^{139-171}$Dy~\cite{Kondev2021}.
Of those, BCPM reproduces the correct $J^{\pi}$ value in six isotopes (35\%), and
in another four (24\%)
the corresponding excited state lies below 300~keV. The quadrupole deformation 
of the isotopes obtained in the BCPM calculation is mostly prolate but there
is a region of oblate ground state in $^{141-153}$Dy.

\subsection{Lanthanum isotopic chain}

\begin{figure}[tb]
\begin{center}
\includegraphics[width=\columnwidth]{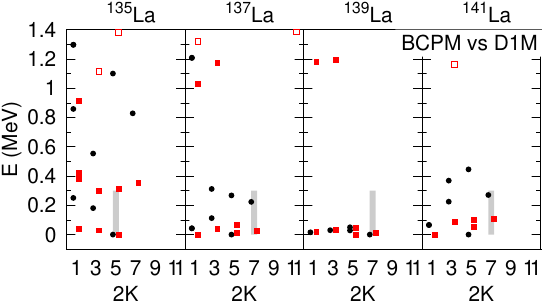}
\caption{The lowest energy spectrum obtained for $^{135-141}$La with the BCPM
functional and the D1M Gogny force using the EFA\@. For an
interpretation of the plot, see caption of Fig.~\ref{fig:Rf-FB-EFA}. In the
present case, black circles (red squares) represent BCPM (D1M) calculations.
Experimental ground state $J$ values are marked with a gray bar. The height of
the bar represents the energy window for the comparison with theoretical results (see text for details).} 
\label{fig:La}
\end{center}
\end{figure}

As an example of odd-$Z$ isotopes, we calculate the low-lying spectra of lanthanum
isotopes. For the odd-$A$ lanthanum isotopes, there are known 13 values of the 
ground-state spin and parity, corresponding to 
$^{125-149}$La~\cite{Kondev2021}. The results of our calculations agree 
with the experiment in four occasions (31\%). This success rate  
is consistent with the 
hit rate obtained with other EDF approaches 
\cite{Bonneau2007a,Afanasjev2011}. If one includes as hits in the 
comparison with experiment those states which in the calculation appear 
at an excitation energy lower than 300~keV, the number of $J^{\pi}$ 
values in agreement goes up to ten (77\% success). 
Of the remaining three cases, two correspond to light proton-rich 
isotopes, and the third to $^{143}$La. The later case deserves further 
theoretical studies considering fluctuations in relevant collective 
degrees of freedom in order to understand the discrepancy. 
The range of isotopes compared include prolate,
deformed, and spherical nuclei. We point out that 
$^{145}$La and $^{147}$La, with neutron numbers close to the ``octupole 
magic number'' $N=88$, are octupole deformed with $\beta_{3} \approx 0.1$. 
The isotope $^{149}$La is also 
octupole deformed, but with a slightly smaller $\beta_{3}$ around 
0.05. However, the depths of the octupole well are only 150, 100, and 
10~keV for $^{145}$La, $^{147}$La, and $^{149}$La, respectively, which are 
not large enough to speak about permanent octupole deformation. 
Given that octupole depth wells tend to become deeper at high spins, it 
is possible for these nuclei to develop alternating parity rotational 
bands. 

In Fig.~\ref{fig:La}, we show the lowest-energy spectra of $^{135-141}$La 
isotopes. There are a bunch of almost degenerate levels in 
$^{139}$La for BCPM, and in $^{137-139}$La for D1M, consequence of a low 
quadrupole deformation compatible with spherical nuclei. Those levels
could be assigned to a spherical $J^{\pi}=7/2^{+}$ ground state in agreement
with experimental data. In the remaining isotopes deformation breaks the
degeneracy but still experimental data are reproduced.

\subsection{Kripton isotopes} 

\begin{figure}[tb]
\begin{center}
\includegraphics[width=0.55\columnwidth]{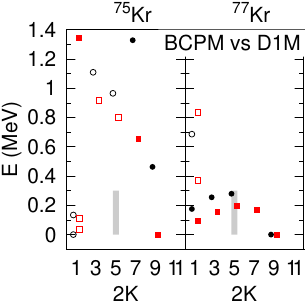}
\caption{The lowest energy spectrum obtained for $^{75,77}$Kr with the BCPM
functional (a) and the D1M Gogny force (b) using the EFA\@. For an
interpretation of the plot, see caption of Fig.~\ref{fig:Rf-FB-EFA}. In the
present case, black circles (red squares) represent BCPM (D1M) calculations.
Experimental ground state $J$ values are marked with a gray bar. The height of
the bar represents the energy window for the comparison with theoretical results (see text for details).} 
\label{fig:Kr}
\end{center}
\end{figure}

Finally, we consider two krypton isotopes as an example of nuclei in 
the low mass region of the nuclear chart. The description of these 
nuclei is very challenging as it is well known that prolate-oblate (and 
triaxial) configurations coexist in a narrow range of energies close to 
the ground state. Therefore, one can consider this comparison as a 
worst case scenario, and  expect a degradation in the comparison 
between BCPM and D1M with respect to the examples discussed in the 
previous sections. In Fig.~\ref{fig:Kr} we show the spectra of 
$^{75-77}$Kr for the two considered interactions. The different levels 
show different values of deformation. For instance, in $^{75}$Kr with 
the BCPM EDF, the $\beta_{2}$ deformation parameters of the five lowest 
states are 0.01, $-0.05$, 0.00, $-0.15$, and 0.04. This alternation of 
spherical, oblate and prolate states is a clear indication of shape 
coexistence. For the same nucleus but with D1M, there is a clear 
dominance of oblate deformations with deformation parameters as large 
as $\beta_{2}=-0.2$, in agreement with the deeper oblate well obtained 
with D1M. For $^{77}$Kr, the deformation obtained with BCPM is almost 
spherical, whereas for D1M it is oblate (but not as large as in 
$^{75}$Kr). Experimental values for the ground state are 
$J^{\pi}=5/2^{+}$ in both cases. The comparison with calculation 
results is not bad for $^{77}$Kr but this is not the case for$^{75}$Kr. 
The differences in deformation discussed above explain the larger 
discrepancies between BCPM, D1M, and experiment found in the present 
case.

\subsection{Multiquasiparticle excitations}

\begin{figure}[tb]
\begin{center}
\includegraphics[width=\columnwidth]{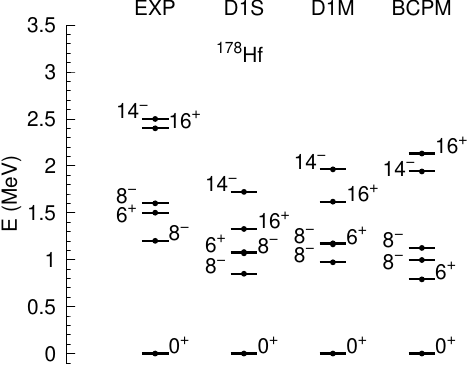}
\caption{High-$K$ states in $^{178}$Hf obtained with the EFA using BCPM, 
Gogny D1S and Gogny D1M are compared with experimental data. The Gogny
excitation energies have been compressed by a factor 0.75 to account for
the effective mass. }
\label{fig:Hf}
\end{center}
\end{figure}

The study and discussion of multiquasiparticle excitations
in atomic nuclei 
is a complex issue that deserves a separate publication.
In this paper, we will just show an example of results obtained for 
high-$K$ excitations in $^{178}$Hf with EFA-BCPM, and compare them with
the homologous results obtained with the D1S and D1M Gogny parametrizations. The choice of
high-$K$ isomers is not arbitrary, as these states are easier to identify experimentally
and less prone to orthogonality issues. In Fig.~\ref{fig:Hf}
the results of EFA calculations for high-$K$ isomeric states in $^{178}$Hf
are shown along with experimental data. There are five known high-$K$ isomers in
this nucleus. The $6^{+}$ is assigned to a two-quasiparticle proton excitation, like
one of the $8^{-}$ state. The other $8^{-}$ is a two-neutrons quasiparticle excitation.
Finally, both the $14^{-}$ and $16^{+}$ states are four-quasiparticle excitations
composed of a two-protons excitation along with a two-neutrons one. One
observes consistency among the theoretical results as well as a
good agreement with experimental data. This example shows that the EFA-BCPM 
provides a suitable description of two and four quasiparticle
excitations.

\section{Conclusions\label{sec:summary}}

We implement the equal filling approximation (EFA) with the BCPM energy
density functional, opening up the calculation of nuclei with an odd
number of protons and/or neutrons. We first test the equivalence between EFA
and full blocking calculations by studying the low-energy spectra predicted by
the Gogny D1S and D1M interaction. We then compare the EFA BCPM predictions
with the D1M ones along several isotopic chains. Once the normalization due to
the effective mass is taken into account, we find that the energy spectra
predicted by the two interactions show a
remarkable similarity, with most of the energy levels agreeing within 200~keV. We
conclude therefore that BCPM and D1M have a similar quality in describing
nuclear low-energy spectra across the nuclear chart. We observe a general
quenching of pairing correlations due to the EFA ``pseudo-blocking'' effect,
which reduces the excitation energy with respect to the unperturbed
one-quasiparticle energy. This result suggests that odd-even staggering in
perturbative calculations could be sensibly reduced in the EFA scheme. We then
performed a comparison with
experimental data for ground state $J^{\pi}$ values, and find that
BCPM can reproduce most of the data within an energy window of 300~keV.
Given the impact of many-body effects beyond mean-field affecting the nuclear
spectrum (such as particle-vibration coupling, symmetry restoration effects,
etc.), we consider the agreement satisfactory, also taking into account that
effective interactions with just a few (tens of) parameters are not able to
capture the full nuclear dynamics across the whole nuclear chart. Finally, two- and
four-quasiparticle excitations are studied by looking at 
high-$K$ isomers in $^{178}$Hf. A close resemblance between BCPM, D1S and
D1M predictions is found, with a sensible reproduction of experimental data. These
results open the door to use BCPM for large scale calculations involving
odd-A and/or multi-quasiparticle excitations.

\begin{acknowledgments}
This work is supported by Spanish Agencia Estatal de Investigacion (AEI) of the
Ministry of Science and Innovation under Grant No. PID2021-127890NB-I00. SG
acknowledges support by the ``Ram{\'o}n y Cajal'' grant No.~RYC2021-031880-I
funded by MCIN/AEI/10.13039/501100011033 and the European
Union-``NextGenerationEU/PRTR''.
\end{acknowledgments}

\input{Paper.bbl}

\end{document}

%% file: Paper.bbl
%